\documentclass[10pt,conference]{IEEEtran}
\IEEEoverridecommandlockouts
\usepackage{cite}
\usepackage{amsmath,amssymb,amsfonts}
\usepackage{algorithmic}
\usepackage{graphicx}
\usepackage{textcomp}
\usepackage{xcolor}
\def\BibTeX{{\rm B\kern-.05em{\sc i\kern-.025em b}\kern-.08em
    T\kern-.1667em\lower.7ex\hbox{E}\kern-.125emX}}
\usepackage[caption=false, font=footnotesize]{subfig}
\usepackage{rotating}
\usepackage{pdflscape}
\usepackage{booktabs}
\usepackage{siunitx}
\usepackage{mathtools}
\usepackage{cuted}
\usepackage{multirow}
\usepackage{eqnarray}
\usepackage{amsmath}
\usepackage{placeins}
\usepackage{float}
\usepackage{lipsum}
\usepackage{flushend}
\usepackage{balance}
\usepackage{scalerel}
\usepackage{stfloats}
\usepackage{bigints}
\usepackage{textcomp}
\usepackage{tabularx}
\usepackage{cleveref}

\begin{document}


\title{{Mitigation of Misalignment Errors \\ Over Inter-Satellite FSO Energy Harvesting}\\
}
\author{\IEEEauthorblockN{Baris Donmez,
Irfan Azam, and
Gunes Karabulut Kurt}
\IEEEauthorblockA{Poly-Grames Research Center, Department of Electrical Engineering,  Polytechnique Montr\'eal, Montr\'eal, QC, Canada\\
Emails: \{baris.donmez,
irfan.azam,
gunes.kurt\}@polymtl.ca}}

\IEEEspecialpapernotice{(Invited Paper)}

\maketitle

\begin{abstract}
In this paper, the impact of the acquisition, tracking, and pointing (ATP) module utilization on inter-satellite energy harvesting is investigated for 1U (0.1$\times$0.1$\times$0.1 m) and 12U (0.2$\times$0.2$\times$0.3 m) satellites for adaptive beam divergence and the corresponding distances while maintaining the spot diameters. Random elevation and azimuth misalignment error angles at both the transmitter and the receiver are modeled with Gaussian distribution hence the radial pointing error angle is modeled with Rayleigh distribution. The Monte Carlo approach is used to determine mean radial error angles for both transmitter and receiver in the non-ATP and ATP cases. The average harvested powers are analyzed as a function of the transmit powers and inter-satellite distances for both 1U and 12U satellites while considering the minimum power requirements. Our simulation results show that in the non-ATP case, the minimum required average harvested power cannot be achieved beyond 680 and 1360 km distances for 1U and 12U satellites, respectively, with a maximum transmit power of 1~kW. However, 2 W of average harvested power can be achieved at around 750 and 1500 km for 1U and 12U satellites, respectively, with a transmit power of 27 W in the presence of an ATP mechanism.  
\end{abstract}


\section{Introduction}
\label{Introduction}

The majority of the small satellites (i.e., CubeSats) operate at 350--700 km altitudes from the Earth's surface. However, there are small satellites including 1U satellites, moving in orbits that have altitudes above 700 km \cite{CubeSatsurvey} (i.e., within the exosphere layer) thus, the losses derived from atmospheric attenuation and scintillation can be neglected \cite{ComprehensivePathLoss}. Solar-powered satellites (SPS) equipped with solar cells are capable to generate sufficient energy from the Sun and the excessive amount can also be used to provide wireless power transmission (WPT) for charging small satellites operating far from the Earth \cite{EHwindow}. Hence, self-sustainability can be achieved in space networks. 

There are different sustainable energy sources other than risky nuclear energy \cite{NuclearSat}. Many current space network applications utilize the microwave radiofrequency (RF) WPT \cite{microSat2Earth} which offers high conversion efficiency and coverage whereas it has some drawbacks as well. The RF WPT has limitations in terms of the maximum achievable range and it is prone to interferences. On the other hand, free space optics (FSO) technology utilizes collimated laser diodes and hence transmits the power on a circular spot area with a smaller spot diameter. Therefore,  the received power by a specific solar cell area is not reduced in FSO WPT contrary to its RF counterpart. However, the main drawback of the FSO WPT systems is misalignment error, and the acquisition, tracking, and pointing (ATP) mechanism must be used for long-range FSO WPT systems (i.e., inter-satellite) to mitigate pointing loss \cite{EHwindow}. Therefore, the random misalignment loss must be considered to compute the harvested power more realistically. 
 
Many existing works on FSO communication systems model the random elevation misalignment error angle and azimuth misalignment error angle statistically with zero-mean, independent, and identically distributed Gaussian distribution. Then, Rician distributed radial error angle model can become Rayleigh distributed when the bias error angle of the Rician distribution is defined as zero \cite{pointingerror,toyoshima_optimum_2002,arnon_beam_1997,Shlomi_optimization_2004}.  

There are a limited number of inter-satellite FSO WPT studies that exist in the literature. In \cite{EHwindow}, the received power and total received energy are presented for the single-hop and cluster scenarios with various distances. In \cite{alouiniEH2}, the harvested power as a function of transmit power for various distances is presented. The authors considered a beam divergence angle of 4 $\mu$rad, and then computed 0.9 kW received power for a 1 kW transmit power over a 25 km line-of-sight (LoS)  distance between an low Earth orbit (LEO) satellite and a CubeSat.  These studies do not consider the losses induced by pointing errors at the transmitter and receiver.      

In our paper, we investigate the adverse effects of random misalignment errors at both the transmitter and the receiver in LoS link between a large SPS and small satellites which have 0.1 and 0.2 m receiver aperture diameters, respectively. The SPS transmits an adaptive collimated laser beam which enables to maintain of an appropriate spot diameter for varying distances and then a smaller satellite harvests the power for operating uninterruptedly since small-size solar arrays cannot generate sufficient energy from the Sun \cite{alouiniEH2}. In a nutshell, we propose a realistic FSO WPT system that considers self-sustainability since the small satellites harvest the power that is transmitted by the SSP satellite. 

The key contributions of this study can be listed as follows:

\begin{itemize}
  \item In our system model, we consider a realistic laser power conversion efficiency (PCE) of a laser diode used in space missions.
  \item We use a realistic energy harvesting conversion efficiency (EHCE) of an appropriate solar cell type.
  \item We generate random misalignment error angles for the transmitter and receiver sides during energy harvesting done by the smaller 1U and 12U satellites.
  \item We investigate the impact of realistic acquisition, tracking, and pointing (ATP) modules for energy harvesting between two satellites.
  \item We compute the maximum inter-satellite distances as 751.88 and 1503.8 km for 1U and 12U satellites, respectively, by adhering to the realistic laser aperture diameter of 8 m.
  \item In the non-ATP case, the 2 W of average harvested power cannot be achieved beyond 680 and 1360 km distances for 1U and 12U satellites, respectively, even with a maximum transmit power of 1 kW. 
  \item We show that the average harvested power of 2 W for CubeSats can be achieved at around 750 and 1500 km for 1U and 12U satellites, respectively, with a transmit power of 27~W in the presence of an ATP mechanism. 
\end{itemize}

The remainder of this paper is organized as follows. In Section \ref{System Model}, we present the system model of the inter-satellite FSO-based energy harvesting system. We elaborate on the misalignment error angle and energy harvesting conversion efficiency model. In Section \ref{Performance Evaluation}, the simulation parameters of our system model are presented and then the performances of our proposed FSO-based energy harvesting system for 1U and 12U satellites with or without the ATP module are evaluated. Finally, we conclude our paper and highlight the future research directions for inter-satellite FSO-based energy harvesting in Section \ref{Conclusions}. 

\begin{figure*}[ht]
\centering
\includegraphics[width=\textwidth]{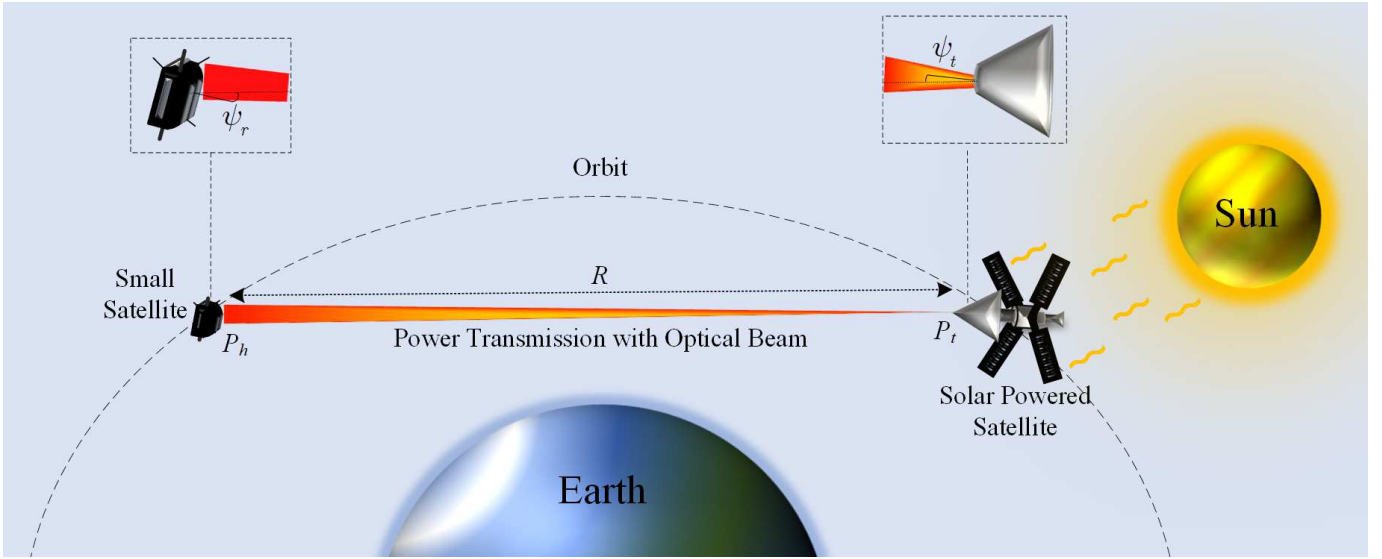}
\caption{System model of inter-satellite optical power transmission and energy harvesting.}
\label{fig1}
\vspace{-0.2cm}
\end{figure*}

\section{System Model}
\label{System Model}

Our proposed system model demonstrated in Fig. \ref{fig1} consists of a larger SPS, and a small satellite (i.e., 1U or 12U) that manages its energy requirement by harvesting energy from the remote laser diode with adaptive beam divergence \cite{adaptivebeamdivergence} that enables to maintain the adequate spot diameter as the distance increases. In our self-sustainable system, we aim to use wavelength-dependent conversion efficiencies as high as possible, however, the efficiency values must be considered for the same wavelengths $\lambda$ for both the transmitter and receiver \cite{EHwindow}. Therefore, the system components must be selected meticulously.   

Efficient laser transmitter selection is a challenging task since the adequate wavelength range for an FSO-based energy harvesting is $\lambda \in (780,1100)$ nm \cite{EHwindow}. For the shorter range FSO links, the beam divergence $\theta$ is in the range of $\theta \in (0.05,1)$ mrad if a tracking module is utilized, otherwise, it is $\theta \in (2,10)$ mrad \cite{OWCmatlab}. However, for inter-satellite links, the required collimated laser beam divergence angles can be determined for a given spot diameter and the LoS distance ${R}$ by using the small-angle approximation as follows \cite{spotdiameter}

\begin{equation}
\theta \text{ }\left[ \text{rad} \right]=\frac{\text{Spot}\,\text{Diameter }\left[ \text{m} \right]}{\text{R }\left[ \text{m} \right]\text{ }}.
\label{EQ:1}
\end{equation}

Moreover, the appropriate aperture diameter of a laser diode, ${{d}_{t}}$, can be determined by ${{d}_{t}}\cong \lambda /\theta $ \cite{OWCmatlab}. The received power of a free space LoS optical link is expressed by \cite{Shlomi_optimization_2004, ComprehensivePathLoss} as follows
\begin{equation}
\small
{{P}_{h}}\!=\!{{P}_{t}}{{\left( \frac{\lambda }{4\pi R} \right)}^{2}}{{\eta }_{e/o}}(\lambda ){{\eta }_{h}}(\lambda){{L}_{t}}({{\psi}_{t}}){{G}_{t}}{{L}_{r}}({{\psi}_{r}}){{G}_{r}}{{L}_{e}}{{L}_{s}}{{L}_{c}},
\label{EQ:2}
\end{equation}

\noindent where ${{P}_{h}}$ is the harvested electrical power, ${{P}_{t}}$ is the transmitted electrical (input) power, ${{\eta }_{e/o}}(\lambda)$ and ${{\eta }_{h}}(\lambda)$ are the wavelength dependant electrical-to-optical PCE and EHCE, respectively. In addition, ${{L}_{t}}({{\psi}_{t}})$ is the radial angle dependant misalignment loss factor at the transmitter, ${{G}_{t}}$ is the transmitter gain, ${{L}_{r}}({{\psi }_{r}})$ denotes the radial angle dependant misalignment loss factor at the receiver, ${{G}_{r}}$ is the receiver gain, ${{L}_{e}}$ is the atmospheric extinction/attenuation loss, ${{L}_{s}}$ is the scintillation loss, and ${{L}_{c}}$ represents the fiber coupling loss. 

\begin{figure}[t!]
\centering
\includegraphics[width=1\columnwidth]{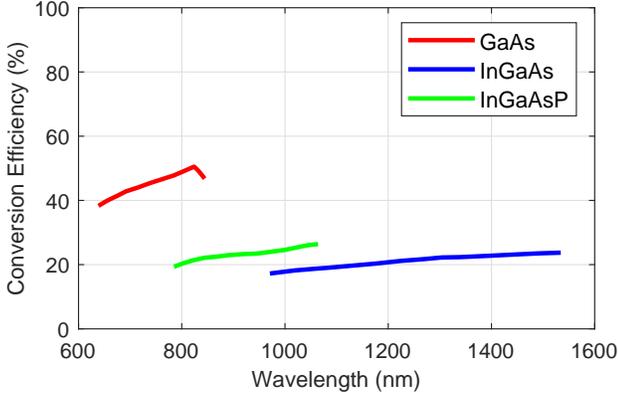}
\caption{Energy harvesting power conversion efficiencies of III-V semiconductor solar cells.}
\label{fig2}
\vspace{-0.2cm}
\end{figure}

As in \cite{ComprehensivePathLoss}, we consider ${{L}_{e}}={{L}_{s}}={{L}_{c}}=1$ for our inter-satellite FSO energy harvesting scenario~\cite{ComprehensivePathLoss}. Furthermore, the transmitter and receiver gains can be approximated as \cite{Shlomi_optimization_2004}
\begin{equation}
{{G}_{t}}\approx {{\left( \frac{\pi {{d}_{t}}}{\lambda } \right)}^{2}},
\label{EQ:3}
\end{equation}
\begin{equation}
{{G}_{r}}\approx {{\left( \frac{\pi {{d}_{r}}}{\lambda } \right)}^{2}},
\label{EQ:4}
\end{equation}

\noindent respectively, where ${d}_{r}$ is the aperture diameter of the receiver.

The misalignment loss factor at the transmitter can be computed by using \cite{Shlomi_optimization_2004}
\begin{equation}
{{L}_{t}}({{\psi }_{t}})=\exp \left( -{{G}_{t}}\psi _{t}^{2} \right),
\label{EQ:5}
\end{equation}

\noindent where ${\psi}_{t}$ is the radial misalignment error angle at the transmitter.

The misalignment loss factor at the receiver is given as \cite{Shlomi_optimization_2004}
\begin{equation}
{{L}_{r}}({{\psi }_{r}})=\exp \left( -{{G}_{r}}\psi _{r}^{2} \right)
\label{EQ:6},
\end{equation}

\noindent that is the radial misalignment error angle at the receiver.

\subsection{Misalignment Error Angle Model}

It is desirable to establish a perfectly aligned LoS optical WPT link throughout the transmission of the laser beam to maximize the received power. The ATP modules enable satellites to mitigate the misalignment errors induced by the mechanical vibrations of the satellites.

The elevation misalignment error angle ${\psi }_{e}$ and azimuth misalignment error angle ${\psi }_{a}$ can be modeled statistically with a zero-mean Gaussian distribution as follows \cite{arnon_beam_1997}

\begin{equation}
f\left( {{\psi }_{e}} \right)=\frac{1}{\sqrt{2\pi \sigma _{e}^{2}}}\exp \left( -\frac{\psi _{e}^{2}}{2\sigma _{e}^{2}} \right)
\label{EQ:7},
\end{equation}

\begin{equation}
f\left( {{\psi }_{a}} \right)=\frac{1}{\sqrt{2\pi \sigma _{a}^{2}}}\exp \left( -\frac{\psi _{a}^{2}}{2\sigma _{a}^{2}} \right)
\label{EQ:8},
\end{equation}

\noindent where $\sigma _{e}^{2}$ and $\sigma _{a}^{2}$ are the variances of elevation and azimuth misalignment angles, respectively. It should be noted that random elevation and azimuth misalignment error angles are independent and identically distributed.

Hence, the radial misalignment error angle ($\psi$) at the transmitter (${\psi }_{t}$) and receiver (${\psi}_{r}$) can be modeled statistically with the Rayleigh distribution by assuming ${{\sigma }_{\psi }}={{\sigma }_{e}}={{\sigma }_{a}}$ due to the symmetry as follows \cite{arnon_beam_1997} 
\begin{equation}
\psi =\sqrt{\psi _{e}^{2}+\psi _{a}^{2}},
\label{EQ:9}
\end{equation}
\begin{equation}
f\left( {{\psi }_{t}} \right)=\frac{{{\psi }_{t}}}{\sigma _{\psi t}^{2}}\exp \left( -\frac{\psi _{t}^{2}}{2\sigma _{\psi t}^{2}} \right),
\label{EQ:10}
\end{equation}
\begin{equation}
f\left( {{\psi }_{r}} \right)=\frac{{{\psi }_{r}}}{\sigma _{\psi r}^{2}}\exp \left( -\frac{\psi _{r}^{2}}{2\sigma _{\psi r}^{2}} \right).
\label{EQ:11}
\end{equation}

\begin{table}[t!]
\centering
\caption{Simulation Parameters}
\label{table1}
\begin{tabular}{|l|l|}
\hline
\multicolumn{1}{|c|}{\textbf{Parameter}} &
  \multicolumn{1}{c|}{\textbf{Value}} \\ \hline
Max. Transmit Power (${P}_{t}$) &
   1 kW \cite{alouiniEH2} \\ \hline
Laser Wavelength (${\lambda}$) & 1064 nm \cite{lasertypes}    \\ \hline
Laser Diode PCE ($\%$)      & 51$\%$ \cite{laser1064}    \\ \hline
\begin{tabular}[c]{@{}l@{}}Beam Divergence Angle ($\theta$) \end{tabular} &
\begin{tabular}[c]{@{}l@{}} Adaptive \end{tabular}  
 \\ \hline
\begin{tabular}[c]{@{}l@{}}Laser Aperture Diameter\\  Constraint (max. $d_{t}$) \end{tabular} &
\begin{tabular}[c]{@{}l@{}} 8 m \cite{maxTxaperdia} \end{tabular}  
 \\ \hline
\begin{tabular}[c]{@{}l@{}}1U Satellite Receiver\\  Aperture Diameter ($d_{r}$) \end{tabular} &
\begin{tabular}[c]{@{}l@{}} 0.1 m \end{tabular}  
 \\ \hline
 \begin{tabular}[c]{@{}l@{}}12U Satellite Receiver\\  Aperture Diameter ($d_{r}$)\end{tabular} &
\begin{tabular}[c]{@{}l@{}}0.2 m \end{tabular}  
 \\ \hline
\begin{tabular}[c]{@{}l@{}}Pointing Resolution \\  \end{tabular} &
  5 $\mu$rad \cite{ComprehensivePathLoss} \\ \hline
\begin{tabular}[c]{@{}l@{}}Pointing Resolution\\ with ATP module \end{tabular} &
0.5 $\mu$rad \cite{nanopointingres} \\ \hline
EHCE ($\%$)                 & 26.4$\%$ \cite{EHCE} \\ \hline
\begin{tabular}[c]{@{}l@{}}Small Satellite Power\\  Requirement (W)\end{tabular} &
  2 W \cite{cubesatPowerReq}  \\ \hline
\end{tabular}%
\end{table}

\subsection{Energy Harvesting Conversion Efficiency Model}
In general, the LoS distance range for inter-satellite connection is considered between 100 m to 250 km \cite{EHwindow}. Hence, although RF and FSO are commonly used technologies for establishing the links between satellites, power transmission by collimated laser offers sufficient energy harvesting at higher LoS distances. Energy harvesting is crucial for smaller satellites since the utilization of solar cell arrays is very limited due to the smaller effective area  \cite{EHwindow}.  

On the other hand, solar cells are preferred to convert signals with frequencies in the spectrum that the sun occupies. Since laser diodes mainly utilize the infrared band such as 1064 nm, solar cells are appropriate for FSO-based energy harvesting~\cite{Solarcells}. 

There are various solar cells converting optical energy to electrical energy, that are made up of different materials. Fig.~\ref{fig2} represents the wavelength-dependent EHCE of some of the common III-V semiconductor solar cells\cite{EHCE}. Although EHCE of GaAs-made solar cells can reach almost 60$\%$ at around 800 nm, it cannot be used straightforwardly since the laser wavelength in our proposed system is 1064 nm as shown in Table \ref{table1}. Therefore, the selection of InGaAsP-made solar cells which offer 26.4$\%$ is suitable for our system.

\begin{figure}[t!]
\centering
\includegraphics[width=\columnwidth]{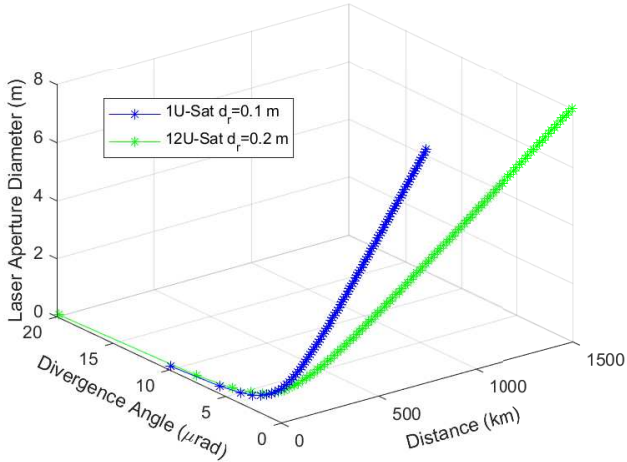}
\caption{Divergence angles and ranges for 1U and 12U small satellites.}
\label{fig3}
\vspace{-0.2cm}
\end{figure}

\section{Performance Evaluation}
\label{Performance Evaluation}

We evaluate the performance of the proposed system through simulations and analyze the impact of ATP on the mitigation of misalignment errors. Before computing the average harvested power, the Monte Carlo method and pointing resolution parameters in Table \ref{table1} are used to analyze independent elevation and azimuth misalignment error angles statistically for both the receiver and transmitter. Then, the mean values of radial pointing error angles are determined for the receiver and transmitter when the ATP module is absent and present. 

\subsection{Simulation Parameters}
\label{Simulation Parameters}
The selection of the proper laser diode type is vital thus, we must consider an operating wavelength that enables sufficient energy harvesting and also can be used in the space applications such as inter-satellite communication or deep-space missions. Hence, the studies of \cite{EHwindow} and \cite{lasertypes} address these concerns, respectively. Therefore, we consider Yd: NVO4 1064 nm laser source with 51\% PCE \cite{laser1064} as mentioned in simulation parameters as given in Table \ref{table1}. 

The adaptive beam divergence angle changes as a function of the spot diameter and distance, and hence the maximum range is determined by considering the maximum laser transmitter aperture diameter as 8 m \cite{maxTxaperdia}. Moreover, we consider various transmit powers ${P}_{t}$ between 1 W and 1 kW in our simulations. 

It should be noted that the EHCE is a wavelength dependant parameter and it must meet the wavelength of the laser source when considering the best possible solar cell type. As per the Fig. \ref{fig2}, InGaAsP offers 26.4$\%$ EHCE which is higher than that of InGaAs.

The 1U and 12U satellites have 0.1$\times$0.1 m and 0.2$\times$0.2 m surfaces thus, collector aperture diameters are considered as $d_r=$ 0.1 and 0.2 m, respectively \cite{NASACubeSatSize}. Besides, the CubeSats require power less than 2 W to accomplish their tasks \cite{cubesatPowerReq}. 

 \begin{figure}[t!]
\centering
\includegraphics[width=\columnwidth]{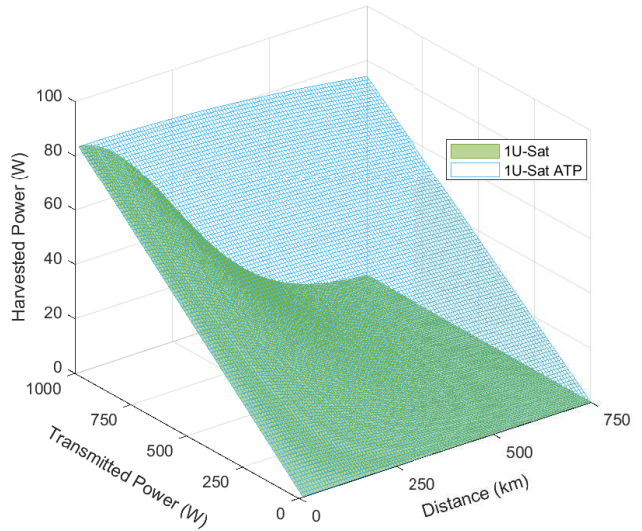}
\caption{Average harvested power for 1U small satellite.}
\label{fig4}
\vspace{-0.2cm}
\end{figure}

\subsection{Results and Discussions}
\label{Results and Discussions}

We have conducted simulations to investigate the impact of ATP with different transmit powers $P_t$ and distances $R$ over the average power harvested $P_h$ by the 1U and 12U satellites in our proposed system model. In addition, we determine the minimum required transmit power to satisfy the power requirement of the small satellites when each small satellite is at the maximum possible distance from the SPS.

By using Eq. \ref{EQ:1}, the adaptive laser beam divergence angles of 1U and 12U satellites are computed for corresponding inter-satellite distances by adhering to the maximum transmitter aperture diameter of 8 m \cite{maxTxaperdia}, as shown in Fig. \ref{fig3}. Due to this laser diode aperture diameter constraint, the maximum achievable distances for 1U satellite with $d_r=$ 0.1 m and 12U satellite with $d_r=$ 0.2 m are 751.88 and 1503.8 km, respectively. Since we consider the minimum inter-satellite distance as 10 km, the widest beam divergence angles for 1U and 12U satellites are 10 and 20 $\mu$rad, respectively, whereas the common narrowest beam divergence angle when $d_t=$ 8~m is 0.133 $\mu$rad.

 For the case of the 1U satellite (0.1$\times$0.1$\times$0.1 m), the improvement made by an ATP mechanism providing a higher pointing resolution  demonstrated by comparing it with the no ATP case in Fig. \ref{fig4}. Since the receiver aperture diameter of 1U satellite is 0.1 m and fixed, the transmitter beam divergence angle adapts itself as the distance increases, to maintain the same spot size. According to Fig. \ref{fig4}, when the distance goes beyond 680 km, the average harvested power drops below 2 W despite the maximum transmit power $P_t=$ 1 kW in the absence of the ATP module. On the other hand, 2 W of average harvested power be achieved even with $P_t=$ 27 W for $R=$~750~km when the ATP module is in use.   

\begin{figure}[t!]
\centering
\includegraphics[width=\columnwidth]{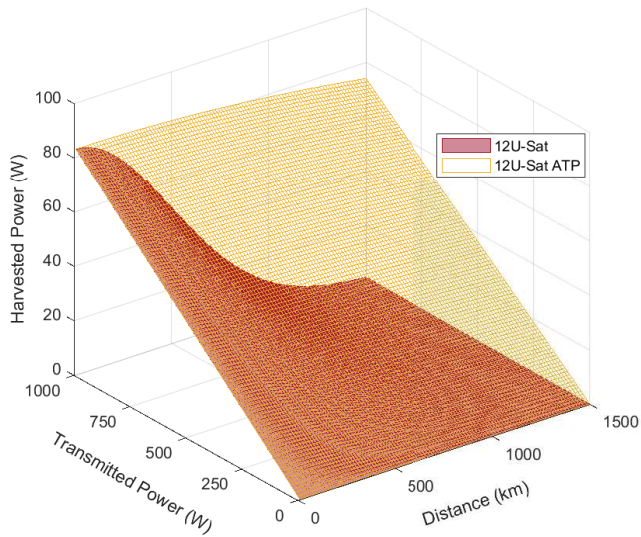}
\caption{Average harvested power for 12U small satellite.}
\label{fig5}
\vspace{-0.2cm}
\end{figure}

 For the case of the 12U satellite (0.2$\times$0.2$\times$0.3 m), the improvement made by the ATP module is presented by comparing it with the no ATP case in Fig. \ref{fig5}. In this case, $d_r=$~0.2~m enables twice the distance for the same laser aperture diameter constraint. Recall that, Eq. \ref{EQ:1} states a beam divergence angle can be maintained despite the distance being doubled if the spot diameter is doubled as well. According to Fig. \ref{fig5}, when $R\ge$ 1360 km and transmit power is maximum as $P_t=$ 1 kW, the average harvested power drops below 2 W without an ATP mechanism. However, when the ATP module is in use, 2 W of average harvested power is achieved with $P_t=$ 27 W for $R=$~1500 km.

\section{Conclusions}
\label{Conclusions}
This paper investigated the role of the tracking module on inter-satellite energy harvesting between 1U and 12U small satellites and the solar-powered satellite which utilizes adaptive beam divergence. Various different transmit powers and adequate ranges were considered as per the laser aperture diameter constraint. Random elevation and azimuth misalignment error angles at both the transmitter and the receiver were modeled with Gaussian distribution. Hence the radial pointing error angle could be modeled with Rayleigh distribution statistically. The narrower the laser beam the lower the misalignment error factor, and hence the average harvested power. 

Our simulation results show a comparison between the energy harvesting made by 1U and 12U small satellites with and without the ATP modules. The outcomes show that ATP is necessary to be able to maximize the inter-satellite distance. For instance,  750 and 1500 km can be achieved for 1U and 12U satellites, respectively, by using $P_t=$ 27 W in the ATP case whereas only around 120 and 240 km can be achieved without the use of the ATP module.  



\bibliographystyle{IEEEtran}
\bibliography{bibliography}

\end{document}